\documentclass[aps,showpacs,prl,toolkits,twocolumn]{revtex4}

\begin{document}


\title {Phase Separation and an upper bound for $\Delta$ for Fermi fluids in the unitary
regime}

\author{Thomas D. Cohen}
\email{cohen@physics.umd.edu}

\affiliation{Department of Physics, University of Maryland,
College Park, MD 20742-4111}

\begin{abstract} An upper bound is derived for $\Delta$ for a cold dilute fluid of
equal amounts of two species of fermion in the unitary regime $k_f a
\rightarrow \infty$ (where $k_f$ is the Fermi momentum and $a$ the
scattering length, and $\Delta$ is a pairing energy: the difference
in energy per particle between adding to the system a macroscopic
number (but infinitesimal fraction) of particles of one species
compared to adding equal numbers of both. The bound is $\delta \leq
\frac{5}{3}\left ( 2 (2 \xi)^{2/5} - (2 \xi) \right)$ where
$\xi=\epsilon/\epsilon_{\rm FG}$ , $\delta= 2 \Delta/\epsilon_{\rm
FG}$; $\epsilon$ is the energy per particle and $\epsilon_{\rm FG}$
is the energy per particle of a noninteracting Fermi gas. If the
bound is saturated, then systems with unequal densities of the two
species will separate spatially into a superfluid phase with equal
numbers of the two species and a normal phase with the excess. If
the bound is not saturated then $\Delta$ is the usual superfluid
gap. If the superfluid gap exceeds the maximum allowed by the
inequality phase separation occurs.
\end{abstract}

\maketitle

During the past several years there has been considerable
theoretical interest in studies of cold, dilute,  Fermi systems with
equal densities of two strongly coupled
species\cite{Ber,Eng,Bak,Hes,Car,Car1,Car2,Asr,Nus,Bay}. In this
context species refers both to spin states and to internal quantum
numbers. There is an idealized version of this problem in which the
following conditions are met: the particles of each species are of
equal mass and $1/a  \ll n^{1/3}\ll 1/r_0$ and $T \ll n^{1/3}$ where
$r_0$ is the typical distance scale of the interaction $a$ is the
scattering length for free particle scattering {\it between} the two
species, $n$ is the density of each species and $T$ is the
temperature and units have been chosen with $\hbar=1$.  In the
extreme limit of this situation where $n^{1/3} a \rightarrow \infty
$, $\Lambda a \rightarrow \infty $, $n^{1/3}/\Lambda \rightarrow 0$
and $T n^{-1/3} \rightarrow 0$, (where $\Lambda$ is the typical
momentum scale characterizing the interaction) there is only a
single momentum scale in the problem, namely $n^{1/3}$.  This is
sometimes called the unitary regime.  It is convenient to re-express
this in terms of a nominal Fermi momentum $k_f \equiv (6 \pi^2
n)^{1/3}$. Thus all physical observables in the problem can be
expressed as appropriate powers of this scale times appropriate
constants. For example the average energy per particle can be
written as $\epsilon=\xi \frac{3}{5} \frac{ k_f^2}{2m}=\xi
\epsilon_{\rm FG}$ where $\xi$ is a universal constant
($\epsilon_{\rm FG}$ is the energy density on a noninteracting Fermi
gas). A pairing energy parameter which gives the difference in
energy per particle between adding a macroscopic number (but
infinitesimal fraction) of particles of one species as compared to
adding equal numbers of both to the system with equal particle
numbers can similarly by given as $2\Delta=\delta
\frac{3}{5}\frac{k_f^2}{2m}=\delta \epsilon_{\rm FG}$.  The
connection between $\Delta$ and the usual superfluid gap is somewhat
subtle and will be discussed below.

This problem is of interest in part due to the universality of the
behavior.  The coefficients such as $\xi$ and $\delta$ apply to {\it
all} problems in this regime regardless of the microscopic details
of the problem.  The problem also is of theoretical interest in that
it represents the exact intermediate limit between two weakly
coupled regimes with $k_f a \ll 1$.  The true weak-coupling regime
between fermions has $k_f a < 0$ and is a BCS superfluid; the regime
with $k_f a$ small and positive corresponds to weakly coupled
molecules in a BEC\cite{Leg}. The problem is also challenging there
appears to be no simple analytical method to compute the universal
coefficients.

The problem is relevant to physical systems of interest. In nuclear
physics the problem of low density neutron matter can be caricatured
by such a system: the two species are the two spin states of the
neutron; the s-wave scattering length between spin up and spin down
neutrons is much larger than the characteristic range of the
nucleon-nucleon force \cite{Bak,nm1,nm2}. The problem has become of
importance at the interface between atomic and condensed matter
since the scattering length between  atoms in particular $m$ states
can be tuned via altering an external magnetic field. The scattering
length diverges at a Feshbach resonance. There has been intense
experimental work on pairing in fluids of trapped Fermionic atoms
and the transition from the BCS to BEC regime\cite{exp}. Of course,
the trap itself can play an important dynamical role in the problem
and significant theoretical effort has gone into describing the role
of the trap which adds a spatial dependence to the
problem\cite{tra}. The present paper will focus be on the ideal case
where the particles are visualized as being contained in a large
box.


The fact that no direct analytical computations of the relevant
dimensionless parameters exists means other methods must be found to
learn something about these dimensionless parameters. One strategy
is attempt to extract them numerically\cite{Car,Car1,Car2,Asr}.  A
possible difficulty with such an approach is that  {\it a priori}
estimates of the errors may be difficult to obtain in a reliable
way. Thus, a constraint based on reliable analytical methods is
potentially quite useful.  One possible idea is to see whether the
coefficients $\xi$ and $\delta$ can be related to each other
analytically. This is is also a formidable challenge for which no
rigorous answer is known. However, as will be discussed in this
paper it is possible to give a rigorous upper bound on $\delta$ for
any assumed value for $\xi$:
\begin{equation}
\delta \leq \frac{5}{3}\left ( 2 (2 \xi)^{2/5} - (2 \xi) \right)
\label{delt}
\end{equation}
As will be discussed below, $\Delta$ need not be the superfluid gap
and the superfluid gap can exceed this bound; if it does one
predicts an interesting phenomenon in the case were the densities of
the two species are unequal.

To derive this bound, consider a generalization of the problem to
the case where the two species (denoted $a$ and $b$) have different
number densities: the only relevant quantities with dimension of
inverse lengths $n_a^{1/3}$ are $n_b^{1/3}$. One completely general
way to parameterize the ground state energy density of this system
subject to the constraint of fixed density of the two species
consistent with the correct dimensional scaling is
\begin{equation}
{\cal E}(n_a,n_b)= \alpha n_a^{5/3} f(n_b/n_a) \label{forma}
\end{equation}
where $f$ is a universal function which only depends on the ratio of
the number densities and $\alpha$ is a constant with dimension of
${\rm mass}^{-1}$.  At the point $n_b=0$ the system is a
noninteracting Fermi gas of species $a$ (by hypothesis the only
relevant interactions are for species $a$ and $b$ to interact with
each other). With out loss of generality one can fix $f(0)$ to be
unity and this in turn fixes $\alpha$ to its Fermi gas value.  The
parametrization in Eq.~(\ref{forma}) is very natural if one
envisions starting with a Fermi gas of species $a$ and slowly adding
in particles of species $b$.  Comparing Eq.~(\ref{forma}) with the
definition of  $\xi$,  one sees that $\xi = f(1)/2$.   The factor of
1/2 in this relation reflects the fact that at $x=1$ the two species
contribute equally; were they non-interacting $f(1)=2$ and $\xi$ is
defined as the fraction relative to the noninteracting case.

The key thermodynamic consideration to derive a bound is  the
possibility of phase separation.  This constrains the energy
density as a function of the densities.  In particular, if we
consider the energy density ${\cal E} (n_a,n_b)$ at two different
pairs of number densities for the two species,
$(n_a^{(1)},n_b^{(1)})$ and $(n_a^{(2)},n_b^{(2)})$, then the
average of the energy densities at these two number densities
cannot exceed the energy density at the average number density:
\begin{widetext}
\begin{equation}
\frac{{\cal E}(n_a^{(1)},n_b^{(1)}) + {\cal
E}(n_a^{(2)},n_b^{(2)})}{2} \geq {\cal E} \left ( \frac{n_a^{(1)}+
n_a^{(2)}}{2},\frac{n_b^{(1)}+ n_b^{(2)}}{2} \right ) \;  ;\label
{av}
\end{equation}
\end{widetext}
if Eq.~(\ref{av}) were false, it would be possible for a system
with a fixed but large volume and number densities $\left (
\frac{n_a^{(1)}+ n_a^{(2)}}{2}, \frac{n_b^{(1)}+
n_b^{(2)}}{2}\right )$ to lower its energy by dividing the volume
into two equal regions with different phases:  one with number
densities $(n_a^{(1)},n_b^{(1)})$ and the other  with
$(n_a^{(2)},n_b^{(2)})$.

Equation \ref{av} implies that in regions where ${\cal
E}(n_a,n_b)$ is continuous its curvature in any direction in the
$n_a,n_b$ plane is positive:
\begin{equation}
\hat{n}_i \frac{\partial^2 {\cal E}}{\partial n_i,\partial n_j}
\hat{n}_j \geq 0 \label{curve}\end{equation}  for all unit vectors
$\hat{n}$ where $i,j$ can assume the value of $a$ or $b$ and
summation of over $i$ and $j$ is implicit.

In order for Eq.~(\ref{curve}) to hold for any unit vector
$\hat{n}$, the matrix
\begin{equation}
{\bf K}(n_a,n_b) \equiv  \left ( \begin{array}{cc}
\frac{\partial^2 {\cal E}}{\partial n_a^2} & \frac{\partial^2
{\cal E}}{\partial n_a
\partial n_b } \\ & \\ \frac{\partial^2 {\cal E}}{\partial n_b
\partial n_a  } & \frac{\partial^2 {\cal E} }{\partial n_b^2} \end{array}
\right ) \label{K}
\end{equation}
must have only nonnegative eigenvalues; thus $\det({\bf K}) \geq
0$.  Inserting the parametrization of Eq.~(\ref{forma}) into the
definition of ${\bf K}$ and imposing a positive determinant yields
a constraint on the curvature of the function $f$
\begin{equation}
f''(x) \geq \frac{ 2 f'^2(x) }{5 f(x)} \label{fcons}
\end{equation}

We know $f(0)=1$ and $f(1)=2 \xi$ and that $f$ continuously connects
these with its curvature constrained by Eq.~(\ref{fcons}).  Consider
the curve which obeys these boundary conditions and saturates
inequality (\ref{fcons}) at all points in between.  Define that
curve as $f_{\rm max}$:
\begin{eqnarray}
f_{\rm max}''(x) & = &
\frac{ 2 {f_{\rm max}'}^2(x) }{5 f_{\rm max}(x)} \
\; \; {\rm where} \; \; =2 \xi \; ;\nonumber \\
{\rm with } \; & \;& \; f_{\rm max}(0)=1 \; \; \; f_{\rm max}(1)
\label{fmaxdef}
\end{eqnarray}
The differential equation in Eq.~(\ref{fmaxdef}) can easily be
solved subject to the boundary conditions.  There is only one real
solution:
\begin{equation}
f_{\rm max}(x) = \left ( 1 + \left ( (2 \xi)^{3/5}  -1 \right ) x
\right )^{5/3} \label{fb}
\end{equation}

The differential equation for $f_{\rm max}$ was derived by
considering a path associated with varying the densities $n_a$ and
$n_b$  which always is in the direction where the second derivative
of ${\cal E}$ is zero: thus the derivatives of ${\cal E}$ with
respect to $n_a$ (or $n_b$) ({\it i.e.} the chemical potentials) are
constants along the path . Therefore $f_{\max}$ represents a
situation in which the system for $x=0$ and $x=1$ and at all points
in between are at the same chemical potential (but different total
density). This is precisely the condition for phase separation: at $
0<x<1$ a fraction $r$ of the particles are in the superfluid phase
with a density of $n_s$ and a fraction $1-r$ are in the normal phase
with a density $n_n$. It is a simple exercise of matching chemical
potentials and densities to show that in such a phase separated
system:
\begin{eqnarray}
n_n & = & n_a \left ( 1 + x ((2 \xi)^{3/5} -1) \right ) \nonumber \\
 n_s & = &
\frac{n_n}{(2 \xi)^{3/5}} \nonumber \\
r & = & \frac{x (2 \xi)^{3/5}}{ 1 + x ((2 \xi)^{3/5} -1) }
\label{phasesep}
\end{eqnarray}
where $n_a$ is {\it average} density of type $a$ over the entire
system.   Note that eqs.~(\ref{fb}) and (\ref{phasesep}) can also be
derived by assuming at the outset two phases and then varying their
densities and fractions subject to the constraints of fixed average
density and fixed ratio of total number of the two species.

The phase separated configuration with fixed $x$ and $n_a$ has a
known energy. The actual minimum energy configuration is either this
energy or below it so $f_{\rm max}$ serves as an upper bound for
$f$:
\begin{equation}
f(x) \leq   f_{\rm max}(x) \label{fbdif} \; ;
\end{equation}
a homogeneous phase violating this condition is energetically
unstable against phase separation and thus the ground state is phase
separated; the upper bound is saturated if phase separation occurs.

It is worth observing that the the preceding analysis is valid only
for $n_b/n_a \leq 1$. However, the regime $x>1$ can easily be
studied as it corresponds to more of species $b$ than $a$. For
$n_b>n_a$ one can use the previous analysis with $b$ and $a$
switched:
\begin{equation}
{\cal E}(n_a,n_b)= \alpha n_b^{5/3} f(n_a/n_b) \; .\label{formb}
\end{equation}
with $f$ the same function as above.

One can determine $\Delta$ from $f(x)$. For an ordinary superfluid
$\Delta$ is the gap. The gap represents the amount of energy saved
by pairing: $ \Delta$ is the difference in energy per particle
gained by adding particles of one species type (say type $a$) to a
system of equal particle number as compared to the energy of adding
equal numbers of $a$ or $b$:
\begin{widetext}
\begin{equation}
2 \Delta=\frac{\left ( E(N+2M,N)-E(N,N) \right )  - \left (
E(N+M,N+M)-E(N,N) \right )}M
\end{equation}
\end{widetext}
where $E(N_a,N_b)$ is the total energy and $N_a$ ( $N_b$) is the
total number of particles of species $a$ ($b$) and $M \ll N$ is
the number of particles of each type added.   Going to the
thermodynamic limit gives $\Delta$ as the discontinuity of the
derivative of ${\cal E}$ with respect to the density of one of the
species:
\begin{equation}
2 \Delta = \lim_{\epsilon \rightarrow 0} \left ( \left .
\frac{\partial{\cal E}}{\partial n_b} \right |_{(nb=na+\epsilon)} -
\left .\frac{\partial{\cal E}}{\partial n_b} \right
|_{(nb=na-\epsilon)} \right )
\end{equation}
Using the general parameterizations of ${\cal E}$ of Eqs.
(\ref{forma}) and (\ref{formb}) yields:
\begin{equation}
2 \Delta =\frac{k_f^2}{2m}\, \xi  \, \left ( 2- \frac{6 f'(1)}{ 5
\xi} \right ) \; \; \;  \delta =  \xi  \, \frac{5}{3} \left ( 2-
\frac{6 f'(1)}{ 5 \xi} \right ) \label{d}
\end{equation}
where the second form follows since $\delta \equiv\Delta/ \left
(\frac{3}{5} \frac{ k_f^2}{2m} \right )$.

Equation (\ref{d}) provides the basis for inequality
(\ref{delt}).  Note that $f \leq f_{\rm max}$ in the interval from
zero to unity and that by construction $f(1)=f_{\rm max}(1)$.
This is possible only if $f'(1) > f_{\rm max}'(1)$.  Thus,
$\delta \leq \xi  \,  \left ( \frac{10}{3}- {2 f_{\rm max}'(1)}/
\xi)\right )$.  Using the explicit form of $f_{\rm max}$ from
Eq.~(\ref{fb}) immediately yields inequality (\ref{delt}).

The interpretation of inequality (\ref{delt}) is subtle. $\Delta$ is
the usual superfluid gap $\Delta_{SF}$ in the case where the system
does not phase separate for unequal numbers ({\it i.e.} the
inequality is not saturated). If there is only one possible phase
when particles are added, then $2\Delta$ must simply represent the
pairing energy for this phase. In the case where inequality is
saturated, however, this is not the case. Although $\Delta$ retains
the definition given above, it should not be interpreted as
$\Delta_{SF}$; if a mixed phase is energetically preferred, $\Delta$
represents the amount of energy per particle to add particles of one
species into a normal phase which forms in equilibrium with the
superfluid phase\cite{San}.  The distinction is the following:
$\Delta_{SF}$ is the energy per particle to a single particle to a
system with equal numbers of the two species; $\Delta$ represents
the energy cost per particle when adding an large number (but
infinitesimal fraction) of particles of one species.
Clearly,$\Delta_{SF} \ge \Delta$ ; either phase separation happens
and the two are equal or it does not and it is energetically cheaper
to add unpaired particles in a new phase .

To summarize: inequality (\ref{delt})always holds with $\Delta$
defined as above.  If in addition the system is known not phase
separate at unequal particle numbers then i) the inequality is not
such saturated and ii) $\Delta=\Delta_{SF}$.

An important corollary of this analysis is that if $\delta_{SF} >
\frac{5}{3}\left ( 2 (2 \xi)^{2/5} - (2 \xi) \right)$ (where
$\delta_{SF}$ is the analog of $\delta$ for the superfluid gap)
phase separation must occur for $x<1$ and if $\delta_{SF} <
\frac{5}{3}\left ( 2 (2 \xi)^{2/5} - (2 \xi) \right)$  then the type
of phase separation considered here (into a fully paired and fully
unpaired phases) does not occur.


Given one nontrivial physical assumption, it is possible to make a
much stronger connection between $\xi$ and $\delta$ than inequality
(\ref{delt}). The dynamical assumption is that for $n_a \neq n_b$,
the system does separate spatially into two phases: a superfluid
phase (which prefers to have equal numbers of the two species) and a
normal phase with the remainder.  This possibility was explored in
an intriguing recent paper by Bedaque, Caldas and Rupak
(BCR)\cite{Bed}. The BCR paper argued on the basis of a generalized
BCS ansatz that that such phase separation occurs. The analysis of
BCR was aimed at a broader class of problems than the strongly
coupled limit of $k_f a \gg 1$; indeed, the detailed analysis is
only strictly legitimate in the case of small BCS gaps and hence
weak coupling.  Thus, it is an open question as to whether such
phase separation occurs at strong coupling and asymmetric systems.
However, BCR argued that it was plausible that their conclusion
holds even away from the weak coupling limit.

Note, that if there is phase separation as suggested by BCR, the
inequalities in (\ref{delt}) and (\ref{fbdif}) must be saturated.
Since the chemical potentials for both species are constant at all
points along the path a region with $n_b=0$ is in chemical
equilibrium a region with $n_b=n_a$; this is necessary and
sufficient for phase separation of the BCR type. Thus, the BCR
assumption implies that for $0> x> 1$, $ f(x) = \left ( 1 + \left (
(2 \xi)^{3/5}  -1 \right ) x \right )^{5/3}$ and thus
\begin{equation}
\delta = \frac{5}{3} \left ( 2 (2 \xi)^{2/5} - (2 \xi)  \right)
\; .\label{delt2}
\end{equation}

Clearly it is important to establish whether phase separation occurs
for $x \ne 1$. As discussed above, this can be immediately answered
if one knows $\xi$ and $\delta_{SF}$ (the analog of $\delta$ for the
superfluid gap).  Estimates of $\xi$ and $\delta_{SF}$ have been
obtained numerically using Monte Carlo methods for finite but large
systems\cite{Car,Car1,Car2,Asr}.  The most recent extracted value
\cite{Car2} of $\xi$ is approximately $.42 \pm .01$ which implies
that the dividing line between whether phase separation occurs or
not is approximately $\delta_{SF}=1.70$ with fairly small numerical
uncertainty.  The extracted value of for $\delta_{SF}$ is $1.68 \pm
.1$. Unfortunately this is not accurate enough to determine whether
phase separation occurs.  The numerical simulations in
ref.~\cite{Car2} for $x<1$, are energetically consistent with phase
separation.  The present analysis implies that if these numerical
simulations reliable, then either $\delta_{SF}$ does exceed
$\frac{5}{3}\left ( 2 (2 \xi)^{2/5} - (2 \xi) \right)$ (presumably
by a small amount) or that phase separation does not occur for $x<1$
with the energy just slightly below the phase separated energy.
Thus, the analysis here provides a highly nontrivial constraint on
the numerics. Finally, it is worth noting that it is surprising just
how close $\delta_{SF}$ is to the critical value for phase
separation: present numerical simulations do not rule out the
intriguing possibility that they are exactly equal.


 \acknowledgments The author acknowledges Shmuel Nussinov and
Boris Gelman for introducing him to the problem.  Sanjay Reddy is
gratefully acknowledged for making an insightful critique about
aspects of previous version of this paper.  This critique clarified
an important point about the nature of the gap. This work was
supported by the U.S.~Department of Energy through grant
DE-FG02-93ER-40762.

\end{document}